\documentclass[12pt,letterpaper]{article}

\usepackage{epsfig}
\usepackage{amsmath}
\usepackage[psamsfonts]{amssymb}
\usepackage{amsthm}
\usepackage{fullpage}
\usepackage{indentfirst}

\numberwithin{equation}{section}

\newcommand{\deltabar}{\overline{\delta}}

\newcommand{\C}{\ensuremath{\mathbb C}}
\newcommand{\PP}{\ensuremath{\mathbb P}}
\newcommand{\Z}{\ensuremath{\mathbb Z}}

\newcommand{\N}{{\mathcal N}}

\newcommand{\M}{{\mathcal M}}
\newcommand{\Mstab}[2]{{\overline{\M_{0,#1,#2}}}}
\newcommand{\Mint}{{\M_{{\rm int}}}}
\newcommand{\Mintgamma}{{\M^\Gamma_{{\rm int}}}}
\newcommand{\Mintlambda}{{\M^\Lambda_{{\rm int}}}}
\newcommand{\Mintone}{{\M^1_{{\rm int}}}}
\newcommand{\Mintgammaprime}{{\M^{\Gamma'}_{{\rm int}}}}
\newcommand{\MintK}{{\M^{K}_{{\rm int}}}}
\newcommand{\Mlines}{{\M_{{\rm lines}}}}
\newcommand{\Mlinesgamma}{{\M^\Gamma_{{\rm lines}}}}
\newcommand{\mulines}{{\mu_{{\rm lines}}}}
\newcommand{\muint}{{\mu_{{\rm int}}}}

\newcommand{\ti}[1]{\textit{#1}}

\newcommand{\CPP}{{\C\PP^{3|4}}}

\newcommand{\prodprime}[2]{{\prod_{#1}^{#2} \!\!\!\!\begin{array}{c}\prime\\ \,\end{array}}}

\def\IC{{\mathbb C}}
\def\A#1{{\cal A}_{[#1]}}        

\def\IA{{\mathbb{A}}}

\def \tilde{\widetilde}

\def \eqref#1{(\ref{#1})}

\def \ignoruj#1{}
\def \eqn#1#2{\begin{equation}#2\label{#1}\end{equation}}

\def \tr{\mbox{Tr\,}}

\def \de{{\rm d}}

\def\ignorethis#1{}
\def\be{\begin{equation}}
\def\ee{\end{equation}}

\def\bear{\begin{eqnarray}}
\def\eear{\end{eqnarray}}

\begin{document}

\bibliographystyle{utphys}

\setcounter{page}{1}
\pagestyle{plain}

\begin{titlepage}

\begin{center}
\hfill HUTP-04/A016\\
\hfill HEP-UK-0021\\
\hfill hep-th/0404085

\vskip 1.5 cm
{\huge \bf Equivalence of twistor prescriptions\\
{\small~\newline} for super Yang-Mills}
\vskip 1.3 cm
{\large Sergei Gukov, Lubo\v{s} Motl, and Andrew Neitzke}\\
\vskip 0.5 cm
{Jefferson Physical Laboratory,
Harvard University,\\
Cambridge, MA 02138, USA}
\vskip 0.3cm
{
{\tt gukov@feynman.harvard.edu}\\
{\tt motl@feynman.harvard.edu}\\
{\tt neitzke@fas.harvard.edu}}
\end{center}

\vskip 0.5 cm
\begin{abstract}
There is evidence that one can compute tree level super Yang-Mills amplitudes
using either connected or completely disconnected curves in twistor space.
We argue that the two computations 
are equivalent, if the integration contours are chosen in a specific way, 
by showing that they can both be reduced to the same integral over a moduli space 
of singular curves.  
We also formulate a class of new ``intermediate'' 
prescriptions to calculate the same amplitudes.
\end{abstract}

\end{titlepage}

\renewcommand{\baselinestretch}{1.4}
\small\normalsize


\tableofcontents

\line(1,0){450}


\section{Introduction}

Recently in \cite{Witten:2003nn} Witten proposed a new approach to
perturbative gauge theories in four dimensions which, among other things,
implies remarkable regularities in the perturbative scattering amplitudes
of $\N=4$ super Yang-Mills and leads to new ways of computing them.
The scattering amplitudes in question depend on the momentum
and polarization vectors of the external gluons,
and are devilishly difficult to compute using the standard Feynman diagram
techniques.
For example, even computing a tree level amplitude with
4 external gluons of positive helicity and 3 gluons of negative helicity
(such an amplitude will be denoted $\A{++++---}$)
requires summing over hundreds of different diagrams!

According to the conjecture of \cite{Witten:2003nn},
perturbative $\N=4$ super Yang-Mills theory can be described
as a string theory in twistor space $\CPP$.
In this reformulation, the Yang-Mills scattering amplitudes
are given by certain integrals over moduli spaces
of holomorphic curves in $\CPP$, which can be interpreted as D1-brane instantons.
More precisely, for a tree level process involving $q$ negative
helicity gluons, the amplitude is given by an integral over
moduli of curves of total degree $d$, where
\eqn{dviaq}{ d=q-1. }
For example, the simplest non-vanishing amplitude with $q=2$
gluons of negative helicity\footnote{We follow the conventions
of \cite{Witten:2003nn} where a $n$-gluon scattering amplitude
is called MHV if $n\!-\!2$ external gluons have positive helicity,
and $\overline{{\rm MHV}}$ (or ``googly'') if $n\!-\!2$ gluons have
negative helicity.} --- the so-called maximally helicity
violating (MHV) amplitude \cite{Parke:1986gb,Mangano:1988xk} --- can be computed by integrating over
the moduli space of degree $1$ curves in $\CPP$ \cite{Witten:2003nn}.

\begin{figure}
\begin{center}
\epsfig{file=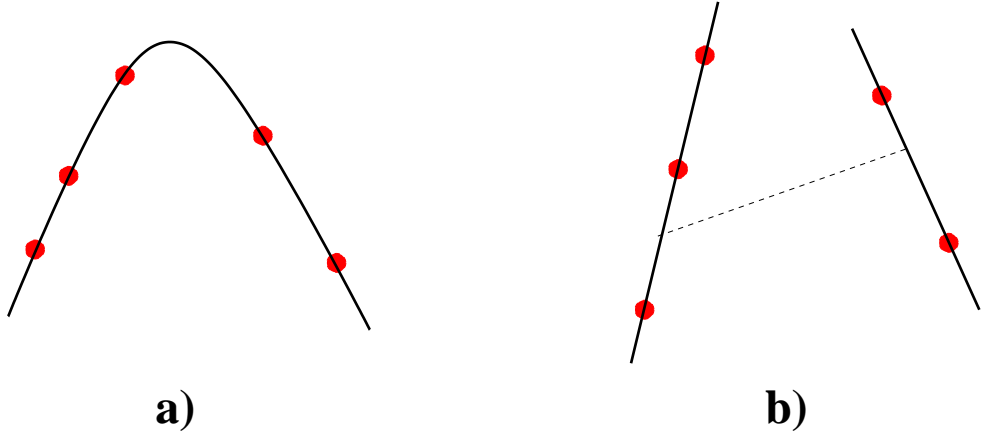,width=130mm}
\end{center}
\caption{An instanton contribution: {\bf (a)} from a connected
curve of degree 2;\quad {\bf (b)} from a pair of degree 1 curves.
The dotted line represents a propagator in holomorphic
Chern-Simons theory.}
\label{linesab}
\end{figure}

However, when one considers the next simplest case, $q=3$, there is a puzzle.
In the prescription of \cite{Witten:2003nn} this amplitude seems to 
involve a sum over two distinct contributions:
one from an integral over connected degree $2$ curves,
and another from an integral over disconnected pairs
of degree $1$ curves; see Figure \ref{linesab}.
Surprisingly, in the case of $\A{++---}$, it was found
that the contribution from connected degree $2$ curves alone
gives the full Yang-Mills amplitude,
at least up to a multiplicative constant \cite{Roiban:2004vt}.
This computation was extended to all googly \cite{Roiban:2004ka}
and some non-MHV \cite{Roiban:2004yf} amplitudes,
again with the surprising result that connected degree $d$ curves already
account for the full Yang-Mills amplitude, without adding any disconnected curves.

On the other hand, there is some evidence that these tree level
amplitudes can also be computed by considering only the contribution
of curves which are ``maximally disconnected,'' namely,
they consist of $d$ distinct degree $1$ lines.
Since degree 1 curves are associated with MHV amplitudes,
this result suggests an alternative method of computing
generic tree amplitudes from graphs with MHV vertices \cite{Cachazo:2004kj}.
The number $v$ of vertices is determined by the number
of gluons with negative helicity; it is actually equal to the degree
\eqref{dviaq},
\eqn{vviaq}{ v=q-1. }
This approach leads to a spectacular simplification of the computations.
For example, the 7-gluon amplitude $\A{++++---}$ mentioned earlier can be
computed using only 8 diagrams with MHV vertices.
However, it also leads to a puzzle.

As we just discussed, the evidence so far in the literature
suggests that rather than one prescription
for Yang-Mills amplitudes there are at least two:
one involving connected curves only, another involving maximally disconnected ones.
We will refer to these as the ``connected prescription''
and the ``disconnected prescription'' respectively.  
These different prescriptions have so far not been related directly.
In a sense, they seem to have complementary virtues:  the connected prescription
expresses the whole amplitude as a single integral, and 
from this form it is easier to prove some properties of the amplitude,
such as the parity symmetry; on the other hand, the disconnected prescription
leads to concrete and immediately useful formulas for the tree level amplitudes.

The purpose of this note is to argue that the connected and disconnected prescriptions
are equivalent, at least for an appropriate choice of the integration contours,
and to give an \ti{a priori} explanation for this agreement.
The explanation is that, in both prescriptions, the integral
over the moduli space is localized to poles on a particular submoduli space. 
This submoduli space parameterizes configurations of intersecting degree $1$ curves.

Let us illustrate this explanation in
the simplest case of degree $2$ curves.
We have two different moduli spaces,  $\Mstab{n}{2}$ and $\Mlines$,
parameterizing respectively connected degree $2$ curves in $\CPP$
and disconnected pairs of lines in $\CPP$,
and integrands $\omega_{{\rm conn}}$
and $\omega_{{\rm disc}}$ on the two spaces (we will review the construction of these integrands
in Section \ref{sec-review}).  Our job is to explain the equality
\begin{equation}
\int_\Mstab{n}{2} \omega_{{\rm conn}} = \int_\Mlines \omega_{{\rm disc}}.
\end{equation}
The explanation begins by noting that both
$\Mlines$ and $\Mstab{n}{2}$ contain a codimension-one
``degeneration locus'' $\Mint$ parameterizing the moduli of
pairs of intersecting lines in $\CPP$.
In the case of $\Mlines$ we get such a degenerate configuration
just by taking two lines in $\CPP$ which happen to intersect.
For $\Mstab{n}{2}$ we get such a degeneration by
considering a hyperbola $xy = C$ in the limit $C \to 0$,
appropriately embedded in $\CPP$.
The crucial point is that both
$\omega_{{\rm conn}}$ and $\omega_{{\rm disc}}$
turn out to have a simple pole along $\Mint$,
and furthermore the residue is the same in both cases.\footnote{We learned of
the possibility of such an explanation from Edward Witten.} Therefore, 
provided that the integration contours on
$\Mlines$ and $\Mstab{n}{2}$ are chosen compatibly
(so that they both encircle $\Mint$ and reduce
to the same contour along it), the desired agreement follows.

The argument for general degree $d$ proceeds along similar lines.  In the moduli space
$\Mstab{n}{d}$ we find a pole where a degree $d$ curve degenerates into two intersecting
curves of degrees $d_1$ and $d_2$; the integral over $\Mstab{n}{d}$ localizes to this sublocus;
then inside this sublocus there is a pole where one of the two curves degenerates further, and
so on until we reduce finally to the moduli space
$\Mint$ of connected trees built from degree $1$ curves.
On the other hand, the integral over $\Mlines$ also reduces to the same $\Mint$, 
because the propagators connecting the different lines have poles when the lines intersect.
Furthermore it turns out that the integrands on $\Mint$ coming from the two prescriptions are
proportional.  This establishes the agreement between these two prescriptions, again provided
that the contours are chosen appropriately, and up to an overall constant which we do not fix.

This iterative argument pays a surprising dividend:  
for any $K = 0, \dots, d-1$, we can define an ``intermediate prescription,''
in which we integrate over configurations of $K+1$ curves with total degree $d$.
We will show that all of these intermediate prescriptions agree with
the connected and disconnected prescriptions.
They can also be understood diagrammatically:
one sums over tree diagrams with $K+1$ vertices,
where each vertex is decorated with a degree.
In these notations,
vertices of degree 1 are the MHV vertices of \cite{Cachazo:2004kj},
whereas vertices with $d>1$ could be called ``non-MHV vertices''.
These intermediate prescriptions deserve further study.

For other recent work on the twistor string approach to Yang-Mills, see
\cite{Roiban:2004vt,Roiban:2004ka,Roiban:2004yf,Berkovits:2004hg,Berkovits:2004tx,Witten:2004cp} 
for the connected prescription, 
\cite{Cachazo:2004kj,Zhu:2004kr,tree-scalar-mhv} for the disconnected prescription,
and \cite{Aganagic:2004yh,Nekrasov:2004js,Neitzke:2004pf} for related
topics.


\subsection{Notation and moduli spaces}

We always consider scattering amplitudes of $n$ external gluons
associated with the particular trace factor $\tr (T_1 T_2 \ldots T_n)$.

We use a coordinate representation
for the super twistor space $\C^{4|4}$.
We unify the bosonic and fermionic indices into
a superspace index $\IA$ taking values in
\eqn{superindex}{\IA\in
\{1,\,2,\,3,\,4\, | 1',2',3',4'\}.}
The components of all objects with bosonic values of the 
superspace index are commuting, while components with fermionic
(primed) values of the superspace index are anticommuting.
The coordinates on the super twistor space will be denoted by $Z^\IA$,
which are related to the coordinates in the literature by
\eqn{twistorcomponents}{(Z^1,Z^2,Z^3,Z^4|Z^{1'},Z^{2'},Z^{3'},Z^{4'})
=(\lambda^1,\lambda^2,\mu^1,\mu^2 |
\psi^{1}, \psi^{2}, \psi^{3}, \psi^{4})
\in {\IC^{4|4}}.} 

We will also be considering various moduli spaces of
curves in $\CPP$  with marked points.  We use the standard notation 
\begin{equation}
\M_{0,n,d}(\CPP)
\end{equation}
for the moduli space of ``genus $0$, $n$-pointed curves of degree $d$ in $\CPP$.''
This moduli space has dimension $(4d+n)|(4d+4)$.
As in \cite{Witten:2003nn} we realize it as the space of
automorphism classes of maps \,$\C\PP^1 \to \CPP$, of degree $d$, with $n$ marked points on $\C\PP^1$.
Since the target space is always $\CPP$ in this paper, 
sometimes we abuse notation and write simply $\M_{0,n,d}$.

\vspace{3mm}

\begin{figure}
\begin{center}
\epsfig{file=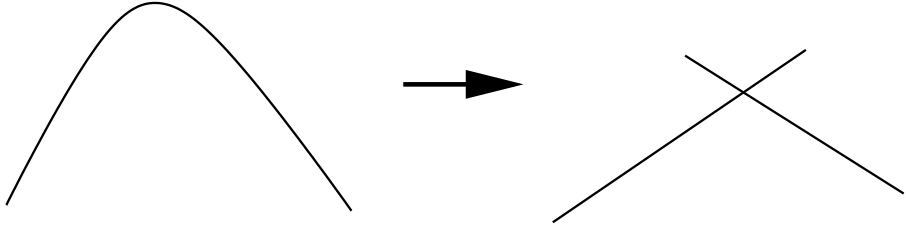,width=120mm}
\end{center}
\caption{A curve of degree 2 can degenerate into a pair
of intersecting lines.}
\label{deg-lines}
\end{figure}

We will be interested in integrating 
over $\M_{0,n,d}(\CPP)$, so we need to understand the
properties of this moduli space.
First, $\M_{0,n,d}(\CPP)$ is non-compact, due
to certain degenerations that a degree $d$ curve with $n$ marked
points can have which are not simply described by a map
$\C\PP^1 \to\CPP$. 
One type of degeneration that will be important
below is when a curve develops a node, {\it i.e.}\ splits into two components.  
There is a standard way of incorporating these degenerate curves into 
our moduli space of maps; one then obtains a larger compact space 
$\overline{\M_{0,n,d}}(\CPP)$, called the ``moduli space of
stable maps.''  This moduli space is a smooth algebraic variety, except for
certain orbifold points which will not play an important role in this paper.\footnote{Strictly
speaking this theorem has been proven when the
target space is $\C\PP^3$ \cite{MR98m:14025}, not for the supermanifold $\CPP$, 
but we do not expect any important differences.}

In particular, the ``boundary'' of this moduli space,
\begin{equation}
\overline{\M_{0,n,d}} (\CPP) \setminus \M_{0,n,d}(\CPP),
\end{equation}
contains a codimension $1$ divisor
which parameterizes curves which have split into two components.
Similarly, for any $K$ there is a subspace $\MintK$
of codimension $K$ that parameterizes reducible curves
with $K$ nodes, {\it i.e.}\ curves which have split up into $K\!+\!1$ intersecting 
components which intersect in a tree.
This $\MintK$ can be further decomposed into irreducible pieces,
\begin{equation}
\Mint = 
\bigcup_\Gamma \Mintgamma,
\end{equation}
where the different $\Gamma$ label 
different shapes of the tree, together with different decompositions
of $d$ into individual degrees $\{d_i\}$, ~
$i=1,2,\dots K\!+\!1$, ~
$d_i \geq d_{i+1}$,\quad and
different ways in which the $n$
marked points can be distributed over the 
$K\!+\!1$
components.
Some of these $\Mintgamma$ will play an important role in our discussion below.


\section{Review of connected and disconnected prescriptions} \label{sec-review}

Suppose we want to use the twistor prescription of \cite{Witten:2003nn}
to evaluate a Yang-Mills amplitude with $q=d+1$ negative helicity gluons.
All contributions to this amplitude are expected to involve holomorphic
curves of total degree $d$, but \ti{a priori} these can be either
connected or disconnected. In this section we review the contributions
which would be expected from the two most extreme cases: connected degree $d$
curves and completely disconnected families of $d$ degree $1$ curves.

In both cases we will consider the Yang-Mills amplitude with arbitrary 
external scattering states.
Via the Penrose transform these scattering states are described by
twistor space wavefunctions,\footnote{Actually,
the wavefunctions are not defined on all of $\CPP$, but this distinction 
will not be important for us.} which are
$\overline{\partial}$-closed $(0,1)$ forms $\phi_i$ ($i=1, \dots, n$) on $\CPP$.  We always treat these $\phi_i$ as generic.
In our computation, we will be focusing on 
poles which arise in integrals over moduli spaces of curves; 
we emphasize that the poles in question never come from the $\phi_i$.

The prescriptions as we write them below are not gauge invariant.
To make the amplitudes gauge invariant we would
probably have to include additional diagrams in both
prescriptions, involving cubic Chern-Simons interaction vertices.
Nevertheless, both prescriptions make sense provided 
we choose a specific gauge for
the gauge field, such as an axial gauge.
In this gauge one expects
that the cubic vertices do not contribute \cite{Witten:2003nn}.\footnote{We
thank Peter Svr\v{c}ek for reminding us of this point.}

\subsection{Connected prescription} \label{review-connected}

We first review the connected prescription for computation
of $n$-point Yang-Mills amplitudes.
The amplitude is obtained as an integral over degree $d$ maps 
\begin{equation}
P: \C\PP^1 \to \CPP.
\end{equation}
Such a map $P$ can be written explicitly, in terms of the inhomogeneous
coordinate $\sigma$ on $\C\PP^1$, as
\eqn{polyno}{
P^\IA(\sigma)=Z^\IA = \sum_{k=0}^d \beta^\IA_k \sigma^k
}
The supermoduli of the degree $d$ map $P$
are $\beta^\IA_k$; these span a space $\C^{4d+4|4d+4}$, which comes
equipped with the natural measure 
\begin{equation}
\mu_d = \prod_{k,\IA} \de \beta^\IA_k.
\label{natural}
\end{equation}
We also have a holomorphic $n$-form on $(\C\PP^1)^n$ given by the free-fermion correlator, 
\begin{equation} \label{free-fermions}
\omega(\sigma_1, \dots, \sigma_n) = 
\prod_{i=1}^n \frac{\de\sigma_i}{\sigma_i - \sigma_{i+1}},\qquad
\sigma_{n+1}\equiv \sigma_1.
\end{equation}

Note that both $\mu$ and $\omega$ are invariant under the group
$GL(2,\C)$ that acts linearly on the homogeneous coordinates on $\C\PP^1$. 
Its action on $\sigma$ is given by the usual expression
\eqn{modular-sigma}{ \sigma \mapsto \sigma' = \frac{a\sigma + b}{c\sigma+ d},
\qquad
ad-bc\neq 0}
while its action on $\beta^\IA_k$ is dictated by the invariance of $Z^\IA$
in \eqref{polyno}: the coefficients $\beta^\IA_k$ may be reorganized 
(up to some combinatorial factors suppressed for simplicity)
into a rank $d$ tensor under $GL(2,\IC)$,
\eqn{ranktensor}{ \{\beta^\IA_k\} = \{\beta^\IA_{I_1 I_2 \dots I_d} \},
\qquad
I_l = 1,2,}
where the number of indices $I_l=2$ equals $k$,
so that the action of $GL(2,\C)$ on $\beta^\IA_k$ becomes
\eqn{action-on-beta}{ \beta^\IA_{I_1 I_2 \dots I_d}
\mapsto {\beta'}^\IA_{I_1 I_2 \dots I_d} =
M_{I_1}^{\,I'_1}
M_{I_2}^{\,I'_2}\dots
M_{I_d}^{\,I'_d}
\beta^\IA_{I'_1 I'_2 \dots I'_d},\quad
M_{I}^{\,I'}=\left(\!\!
\begin{array}{rr}d&-b\\ -c&a\end{array}
\right)
.}
Along with $\mu$ and $\omega$ we also have to include the external wavefunctions,\footnote{We write
$\phi(P(\sigma_i))$ for the pullback of $\phi$ to moduli space via
the evaluation map sending $P$ to $P(\sigma_i)$.}
\begin{equation}
\Phi = \prod_{i=1}^n \phi_i(P(\sigma_i)).
\end{equation}
Putting everything together,
the Yang-Mills amplitude is formally\footnote{Here and below,
by $\mathrm{vol}(GL(2,\C))$ we really mean the volume form
on that group; this is just the standard quotient, when written
in terms of an integral over the quotient space.}
\begin{equation} \label{formal-yma}
\int_{\Mstab{n}{d}} 
\frac{\mu_d \wedge \omega(\sigma_1, \dots, \sigma_n)}{\mathrm{vol}(GL(2,\C))}
\wedge \Phi.
\end{equation}

The expression \eqref{formal-yma} is formal for several reasons.
The first and most serious reason is that we have to choose a contour
for the integral over the coordinates $\beta^\IA_k$ in 
$\Mstab{n}{d}$, and the proper choice of contour
is not yet well understood. (We do not have
to choose a contour for the integrals over $\sigma$,
because the integrand includes both $\de \sigma$ from $\omega$
and $\de \bar{\sigma}$ from the external wavefunctions.)
We will have more to say about the contour below; to match the
disconnected prescription we will 
essentially use a contour around infinity (suitably defined) so that {\it all}
residues are counted.

Second, we have to divide out by the action of $GL(2,\C)$.
A convenient gauge-fixing will be chosen below, but of course the amplitude
is independent of the choice of gauge.
We should perhaps mention that we consider $GL(2,\C)$ over $\C$, {\it i.e.}\  
we divide by the ``holomorphic'' volume form.  This means that
\begin{itemize}
\item this symmetry will always be fixed by a set of holomorphic
conditions; 
\item we will sum over all inequivalent solutions;
\item only the holomorphic Jacobian will be included in the integrals.
\end{itemize}
These rules are compatible with the computations of
\cite{Roiban:2004vt,Roiban:2004ka,Roiban:2004yf}.


\vspace{3mm}

\begin{figure}[t]
\begin{center}
\epsfig{file=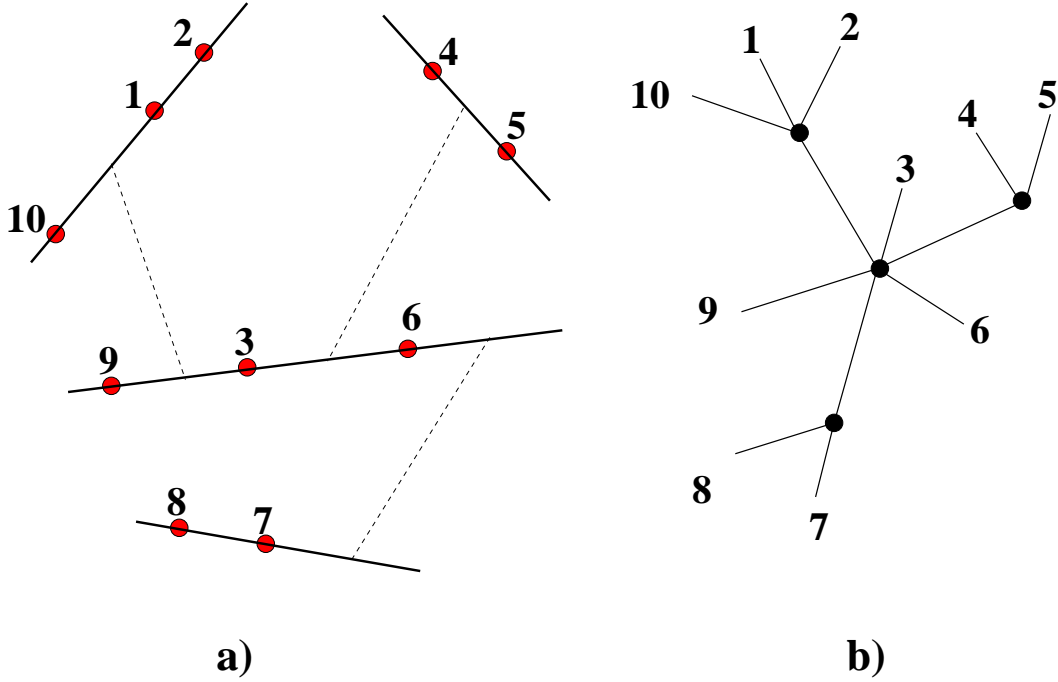,width=140mm}
\end{center}
\caption{A contribution to Yang-Mills amplitudes with $5$ positive
and $5$ negative helicity gluons, represented
{\bf (a)} as four disconnected lines in twistor space,
{\bf (b)} as a graph $\Gamma$ with four MHV vertices.
}
\label{fig-mhv-tree}
\end{figure}

\subsection{Disconnected prescription} \label{review-disconnected}

Now we describe the disconnected prescription for the same amplitudes, 
formulated in twistor space along the lines of
the derivation given in \cite{Cachazo:2004kj}.
In this prescription a tree level amplitude involving $d+1$ negative helicity
gluons, with a particular cyclic ordering, is obtained as a sum over
various tree diagrams with $d$ vertices.
In Figure \ref{fig-mhv-tree} we show a representative example of a 
diagram $\Gamma$ which contributes to
amplitudes with $5$ positive and $5$ negative helicity gluons.
The $10$ external gluons are arranged cyclically around the index loop,
and since there are $5$ negative helicity gluons there are $5-1=4$ vertices.
The vertices have arbitrary
valence.\footnote{Ultimately, it turns out that any diagram containing
a vertex of valence $\le 2$ does not contribute to the amplitude \cite{Cachazo:2004kj}.}
We have not specified which gluons have which helicities;
the twistor space computation yields superspace expressions
which generate the answers for all possible choices
when suitably expanded in the fermionic coordinates.

\vspace{3mm}

\begin{figure}
\begin{center}
\epsfig{file=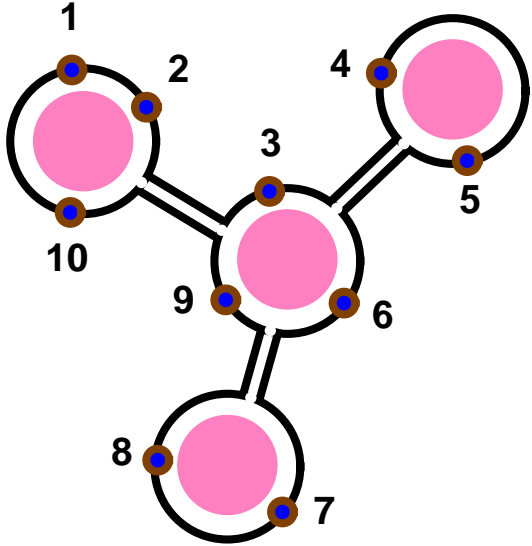,width=70mm}
\end{center}
\caption{A different version of Figure \ref{fig-mhv-tree},
representing the same single-trace amplitude
with the index line made manifest. The circles represent
degree 1 curves in twistor space.}
\label{fig-mhv-tree-index-line}
\end{figure}

Each vertex of $\Gamma$ corresponds to a $\C\PP^1$ in $\CPP$, equipped with
marked points corresponding to internal or external lines attached to the vertex.
To compute the contribution of $\Gamma$ to the amplitude we have
to integrate over the moduli of these curves,
given by $d$ degree $1$ maps
\begin{equation}
Q_i: \C\PP^1 \to \CPP.
\end{equation}
Each such map can be written
\begin{equation}
Q_i^\IA(\sigma) = \sum_{k=0}^1 \beta^\IA_{k,i} \sigma^k
\end{equation}
so there are a total of $8d|8d$ supermoduli $\beta^\IA_{k,i}$ for these $d$ maps,
reduced to $4d|8d$ by the $GL(2,\C)^d$ symmetry.  We also have to integrate over the
moduli for the marked points; if in the diagram $\Gamma$ there
are $n_i$ marked points on the $i$-th $\C\PP^1$, 
then the full moduli space is
\begin{equation}
\Mlinesgamma = \prod_{i=1}^d \M_{0,n_i,1}(\CPP). 
\end{equation}

As in the connected case there is a natural measure
for the moduli of the curves,
\begin{equation}
\mulines = \prod_{k,\IA,i} \de \beta^\IA_{k,i}.
\end{equation}
There are several factors in the integrand which depend on the marked points.
First, there is a free-fermion correlator for each curve; the points on the $i$-th
$\C\PP^1$ come with a cyclic ordering as indicated in Figure \ref{fig-mhv-tree}, and if
we label them $\sigma_1, \dots, \sigma_{n_i}$, they contribute
\begin{equation}
\omega_i = \omega(\sigma_1, \dots, \sigma_{n_i})
\end{equation}
with $\omega$ defined in \eqref{free-fermions}.
These free-fermion correlators contain $\de \sigma$ for each
marked point.

Next we have to include the external wavefunctions:
each external wavefunction $\phi_j$ is connected
to a marked point $\sigma$ on the $i$-th $\C\PP^1$,
for some $i$, and the integrand includes the factor 
\begin{equation} \label{external}
\phi_j(Q_i(\sigma))
\end{equation}
just as in the connected prescription.
But unlike the connected prescription, here we also have some marked points
which are connected to internal propagators.
Let us write $D(\cdot, \cdot)$ for the twistor space propagator,
which is a $(0,2)$-form on $\CPP \times \CPP$.
Each internal propagator is connected to two
marked points $\sigma$, $\sigma'$ on the $i$-th
and $i'$-th $\C\PP^1$'s respectively,
for some $i$, $i'$, and contributes to the integrand a factor
\begin{equation} \label{internal}
D(Q_i(\sigma), Q_{i'}(\sigma')).
\end{equation}
Let us write $\Phi \wedge D$ for the product of all the wavefunctions
and propagators from \eqref{external}, \eqref{internal}.
Since every marked point is attached either
to a propagator or to an external wavefunction, this $\Phi \wedge D$
includes one factor $\de \bar{\sigma}$ for each marked point.

Then the amplitude in the disconnected prescription is
given by the sum over tree diagrams,
\begin{equation}
\sum_{\Gamma} \int_{\Mlinesgamma} 
\frac{\mulines
\wedge \left(\prod_{i=1}^d \omega_i \right) 
\wedge \Phi\wedge D}{\mathrm{vol}(GL(2,\IC))^d}
\label{discpr}.
\end{equation}

As with the connected prescription, to make this integral
concrete we have to do two more things.
First, we must gauge-fix the symmetry $GL(2,\C)^d$ which
acts separately on each $\C\PP^1$.  Second, we must choose
a contour for the integrals over the moduli $\beta^\IA_{k,i}$.

In \cite{Cachazo:2004kj} it was argued that if one makes a particular choice
of contour, and chooses external wavefunctions corresponding to gluons of fixed helicity
and momentum, then the integral over $\Mlinesgamma$ in \eqref{discpr} can be evaluated by a simple rule.
Namely, one first assigns $(+)$ and 
$(-)$ helicities to the endpoints of each propagator,
consistent with the rule that each vertex should have exactly two $(-)$ helicities on it;
for given $\Gamma$, 
there is at most one way to do this.  (If there is no way to do it, then the diagram $\Gamma$
just contributes zero.)  Then each vertex gives a copy of the MHV 
amplitude --- continued off-shell in a specific way to accommodate the internal lines ---
while each propagator carrying momentum $q$ gives $1/q^2$.

For future use in section \ref{csw-generalized} we also mention a natural
generalization of the disconnected prescription:  instead of using $d$ degree $1$ curves
we could use $K+1$ curves for some $K$, with total degree $d$, 
connected into a tree by $K$ propagators.  The integrand is then defined in a way
precisely analogous to \eqref{discpr}, except that the sum over $\Gamma$ includes 
all choices for the degrees of the curves in addition to distributions of the marked points.


\section{Matching the prescriptions in degree $2$ case}

\subsection{The argument in degree 2 case}

How can the disconnected and connected prescriptions give the same result?
Let us consider next-to-maximally helicity violating amplitudes, $q=3$,
which come from degree $2$ curves.  We postpone the discussion
of curves of higher degree to section 
\ref{higherdegree}.

The contribution of disconnected
instantons comes from pairs of degree $1$ curves connected by a
single propagator, with $n$ marked points distributed over the pair of
curves.  This moduli space has dimension $(8+n)|16$ (which includes
$4|8$ for each degree $1$ curve
plus $n$ for the marked points.)
Different distributions of the marked points correspond to
different MHV diagrams $\Gamma$.\footnote{There 
are $n(n+1)/2$ such diagrams, although once
we fix the external wavefunctions not every diagram gives a nonzero contribution
to the sum \eqref{discpr}; 
if the helicities are $---+++ \cdots ++$, then there are $2(n-3)$
diagrams which contribute.}

\begin{figure}[t]
\begin{center}
\epsfig{file=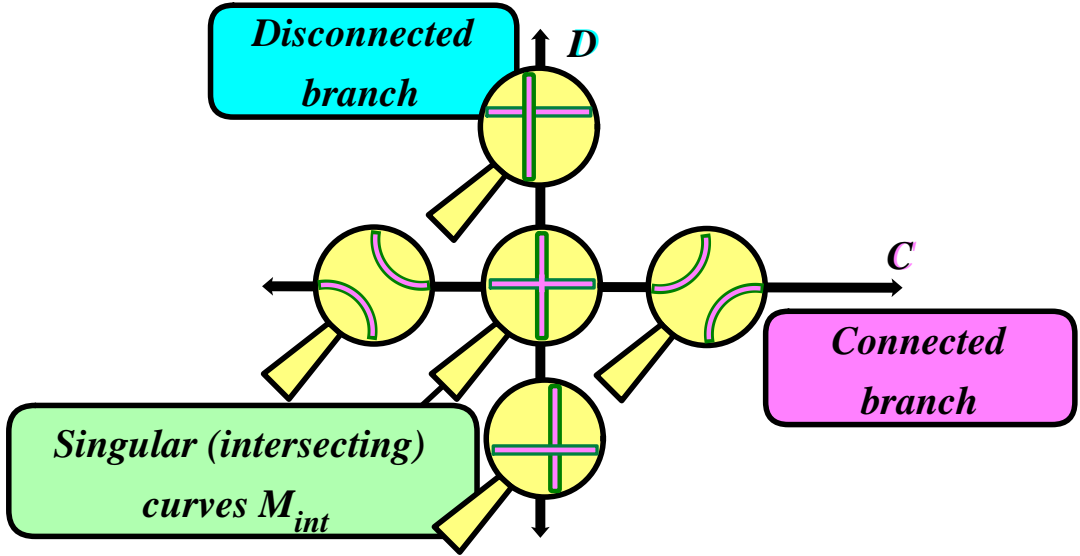,width=145mm}
\end{center}
\caption{A degenerate configuration of two intersecting lines
in $\CPP$ can be deformed into a smooth connected curve
of degree 2 or into two disconnected lines.
The transition between the two branches of
moduli space is reminiscent of
a conifold transition.
}
\label{branches}
\end{figure}

It was shown in \cite{Cachazo:2004kj} that for each $\Gamma$ the
integrand in \eqref{discpr} has a simple pole on the submoduli space $\Mintgamma$,
parameterizing degenerate configurations of intersecting lines of degree 1.  
This submoduli space has dimension $(7+n)|12$, because
the condition that there exists an intersection in the bosonic space 
removes one bosonic modulus, and the condition that all four fermionic 
coordinates of the two lines coincide at this point removes four fermionic 
moduli.\footnote{This fermionic delta-function guarantees the opposite helicity of
the two endpoints of the propagators when one expands in fermions to evaluate a 
particular amplitude.}

After contour-integrating to localize to $\Mintgamma$, the sum \eqref{discpr} can be written as
\eqn{dtwoi}{
\sum_{\Gamma} \int_{\Mintgamma} 
\frac{1}{\mathrm{vol}(GL(2,\C))^{2}}
\left( \muint \wedge
\left( \prod_{i=1}^{n_1} \frac{\de\sigma_i}{\sigma_i - \sigma_{i+1}} \right)
\wedge
\left( \prod_{j=1}^{n_2} \frac{\de\sigma_j'}{\sigma_j' - \sigma_{j+1}'} \right) \right)
\wedge \Phi.
}
Here $i$ and $j$ run over the marked points on each $\C\PP^1$,
including the point of intersection;
so for a diagram with $m$ external wavefunctions
attached to the first line, $n_1 = m + 1$
and $n_2 = n - m + 1$.  
Also, $\sigma_{n_1+1} \equiv \sigma_1$ and $\sigma'_{n_2+1} \equiv \sigma'_1$.
The measure $\muint$ is completely determined by the symmetries
of $\CPP$.

On the other hand, from the connected
prescription \eqref{formal-yma} we find
\eqn{dtwoc}{
\int_{\Mstab{n}{2}}
\frac{1}{\mathrm{vol}(GL(2,\C))}
\left( \mu_2 \wedge
\left( \prod_{i=1}^{n} \frac{\de\sigma_i}{\sigma_i - \sigma_{i+1}} \right) \right)
\wedge 
\Phi.
}
We will reorganize the integral \eqref{dtwoc} over the $(8+n)|12$-dimensional
space $\Mstab{n}{2}$ of conics in the following way:
Locally, to any conic we will associate a pair of intersecting lines which 
are its ``asymptotes.''  The moduli space of pairs of intersecting lines 
with $n$ marked points is  
the $\Mint$ which occurred in the disconnected prescription.  
This $\Mint$ has 
dimension $(7+n)|12$, so in $\Mstab{n}{2}$ there is one more coordinate, which we call $C$;
$C=0$ corresponds to the singular conics, which coincide with their
asymptotes. 
This $C$ can be thought of as a ``deformation parameter'' which resolves the singularity.
We will find that the integrand has a pole at $C=0$, {\it i.e.}\ along
$\Mint$.

More precisely, $\Mint$ includes only those degenerations in which the marked
points are distributed 
in a way corresponding to some MHV tree graph $\Gamma$.  This just means the points are broken
into two groups which are cyclically ordered --- so e.g.\ if $n=6$, there is a component of $\Mint$ with points $1,2,3$ 
on one line and $4,5,6$ on the other, but we do not include the degeneration which has $1,2,4$ on one line and $3,5,6$ on the other.  Indeed, we will see that the latter degeneration 
does \ti{not} give a pole.
We will find poles only along $n(n+1)/2$ distinct components $\Mintgamma$, 
which are in one-to-one correspondence with the diagrams $\Gamma$ contributing to \eqref{dtwoi}.

Moreover, we will show that the residue along $\Mintgamma$ is precisely
such that the integral \eqref{dtwoc} agrees with \eqref{dtwoi} after localizing.
This will complete the argument for the equivalence in the degree $2$ case.

\subsection{Computing the residue in degree 2 case\label{lower-residue}}

In this section we show that the integral \eqref{dtwoc} over the moduli
space $\Mstab{n}{2}$ of genus zero, degree 2 curves in $\CPP$ with $n$ marked points 
has a pole at the subspace $\Mint$ describing
pairs of intersecting lines, and that
it has the desired residue as discussed in the last section.

Let us start by fixing part of the $GL(2,\C)$ symmetry reviewed in section
\ref{review-connected}.
We use three generators of $GL(2,\C)$ to impose
the constraints
\eqn{gaugechoice}{
P^4(\sigma) = \sigma
\mbox{\qquad {\it i.e.}\qquad}
(\beta^4_0,\beta^4_1,\beta^4_2) = (0,1,0).
}
In other words, we are imposing the conditions that the two intersections 
of the hyperplane $Z^4\!=\!0$\, with the curve have
coordinates\footnote{The point $\sigma=\infty$ can be written
as $(1:0)$ in homogeneous coordinates, and therefore is 
completely nonsingular.}
$\sigma\!=\!0$ and $\sigma\!=\!\infty$, and normalizing
the coefficients $\beta^4_{0,1,2}$.
There is one more generator of $GL(2,\C)$
to be fixed, the matrix
\begin{equation}
M = \begin{pmatrix}
\lambda & 0 \\
0 & \lambda^{-1}
\end{pmatrix},
\end{equation}
which acts as
\eqn{unfixed}{\beta^\IA_k \to \lambda^{2-2k} \beta^\IA_k,
\qquad
\sigma\to \lambda^2 \sigma.
}
This transformation preserves the gauge choice \eqref{gaugechoice}.


\subsubsection{Factors from the measure on the moduli space}

Using the freedom to divide all twistor coordinates $Z^\IA$ by $\sigma$,
we can write \eqref{polyno} as
\eqn{polynoo}{
P^\IA(\sigma)=Z^\IA = \sum_{k=0}^{2} \beta^\IA_{k} \sigma^{k-1} =
\beta^\IA_0\sigma^{-1} + \beta^\IA_1 + \beta^\IA_2 \sigma,
}
which using \eqref{gaugechoice} implies $P^4(\sigma)=1$.  As $\sigma \to \infty$ or $\sigma \to 0$,
we can neglect the first or the last term in \eqref{polynoo}, respectively.
So \eqref{polynoo} describes a hyperbola that approaches
two asymptotic lines in the superspace $\IC^{3|4}$:
\eqn{asymplines}{Z^\IA=
\beta^\IA_0\sigma^{-1} + \beta^\IA_1,
\qquad
Z^\IA = \beta^\IA_1 + \beta^\IA_2 \sigma.
}
These two lines intersect at the point $Z^\IA=\beta^\IA_1$, while
$\beta^\IA_0$ and $\beta^\IA_2$ with $\IA\neq 4$
are the tangent vectors along 
these lines.  It is important that for every conic
$\Sigma := P_* (\C\PP^1) \subset \CPP$ we can find a singular
conic $\Sigma'$ (a pair of intersecting lines) in $\Mint$
defining the asymptotes of $\Sigma$.
This rule is not canonical; 
it depended on our choice to single out 
the points at infinity, {\it i.e.}\ the hyperplane $Z^4\!=\!0$.

We want to express $\Mstab{n}{2}$ locally
as a product of $\Mint$ and $\C$, with the extra $\C$ parameterized
by the deformation parameter $C$.  What are the appropriate coordinates?
The $3|4$ parameters 
\eqn{firstmoduli}{\beta^\IA_1,\qquad \IA\neq 4,} 
describing the position of the intersection of the asymptotes,  
give coordinates on $\Mint$.  The remaining $4|8$ 
coordinates on $\Mint$ are the directions of the two asymptotes; 
each asymptote gives us $2|4$ moduli.  
We want to describe these directions by ``unit vectors'' in a suitable sense.
As we approach a generic point of $\Mint$, 
$\beta^3_0$ and $\beta^3_2$ are nonzero, and we 
may use them to normalize the direction vectors. 
In other words, the remaining $2|4$ plus $2|4$ coordinates on $\M_{{\rm int}}$ may be 
chosen as
\eqn{remainingm}{
\frac{\beta^\IA_0}{\beta^3_0}
\mbox{\quad and \quad}
\frac{\beta^\IA_2}{\beta^3_2},\qquad
\IA\in\{1,\,2\,|1',2',3',4'\}.
}
(Choosing different coordinates on $\Mint$ instead of \eqref{firstmoduli}
and \eqref{remainingm} would not change the result below; the only change 
would be a $C$-independent Jacobian.)

Looking at our original coordinates on $\Mstab{n}{2}$, we still have
two more bosonic components of $\beta$ which are independent of our
coordinates on $\Mint$, namely $\beta^3_0$ and $\beta^3_2$ themselves.
We also have one unfixed generator of $GL(2,\C)$ given in
\eqref{unfixed}.  This generator simply multiplies the ratio 
$\beta^3_0 / \beta^3_2$ by $\lambda^4$, so we can use it to fix that
ratio to a constant, such as
\eqn{gaugechoicee}{\frac{\beta^3_0 }{\beta^3_2} = 1.} 
Now having fixed the full $GL(2,\C)$ symmetry we can write the measure $\mu_2$
from \eqref{natural} as
\begin{equation} \label{measure-deltas}
(J/4)\ \prod_{k, \IA} \de \beta_k^\IA\ \delta(\beta_0^3 / \beta_2^3 -1)\ \delta(\beta_0^4)\ \delta(\beta_1^4 - 1)\ \delta(\beta_2^4).
\end{equation}
Here $J$ is the determinant of the Jacobian matrix of variations of the constraints
with respect to the $GL(2,\C)$ generators.
If we parameterize the generators of $GL(2,\C)$ by
\begin{equation}
M = \begin{pmatrix}1+a & b \\ c & 1+d\end{pmatrix}
\end{equation}
then this matrix is
\eqn{fixingmatrix}{
\delta
\left(
\begin{array}{c}
\beta^4_0\\
\beta^4_1\\
\beta^4_2\\
\beta^3_0 / \beta^3_2
\end{array}
\right)
=
\left(
\begin{array}{rrrr}
\!\!1&0&0&0\\
0&1&0&0\\
0&0&1&1\\
*&\phantom{-}*&\phantom{-}2&-2
\end{array}
\right)
\left(
\begin{array}{c}
b\\
c\\
a\\
d
\end{array}
\right)
}
and hence we get simply
\begin{equation}
J = -4.
\end{equation}
The factor $J/4$ in \eqref{measure-deltas} represents
$1/\mathrm{vol}(GL(2,\C))$; we had to divide by $4$
because the $\Z_4 \subset GL(2,\C)$ generated by
\begin{equation} \label{unfixed-discrete-matrix}
M = 
\left(\begin{array}{rr}
  i    & 0 \\
  0    & -i
\end{array}\right)
\end{equation}
is left unfixed by our gauge condition.

The three delta functions in \eqref{measure-deltas} involving
$\beta_k^4$ just eliminate the integrals over those variables,
imposing \eqref{gaugechoice}.  Let us
also use $\delta(\beta_0^3 / \beta_2^3 -1)$ to eliminate $\beta_0^3$,
imposing \eqref{gaugechoicee}.
Integrating over $\beta_0^3$ gives
a factor $\beta_2^3$, so the measure becomes
\eqn{natura}{
- \beta_2^3\,\de \beta_2^3
\prod_{\IA\neq 4} \de \beta_1^\IA
\prod_{k\in\{0,2\}}  \prod_{\IA\neq 3,4}
\de
\beta_k^\IA.
}
We rewrite this as a measure for
the single transverse coordinate $\beta_2^3$, times
a measure on $\Mint$, for which a full set of $7|12$ coordinates were given 
in \eqref{firstmoduli}, \eqref{remainingm}:
\eqn{naturab}{
\left(- (\beta_2^3)^{1-4} \de \beta_2^3\right)
\times \left( \prod_{\IA\neq 4} \de \beta_1^\IA
\prod_{k\in\{0,2\}}  \prod_{\IA\neq 3,4}
\de
\left(
\frac{\beta_k^\IA}{\beta_k^3}
\right) \right).
}
The extra power $(-4)$ in $(\beta_2^3)^{-4}$ was calculated as
$2_{k=0,2}\times (2_B-4_F)$; the terms $2_B$ and $-4_F$ arise from the
redefined bosonic {\it and} 
fermionic measures involving $\beta_k^\IA$, respectively.

The coordinate $\beta_2^3$ is related to 
the deformation parameter $C$ --- we will see
that the natural definition of $C$ is $(\beta_2^3)^{2}$.
The measure
$(\beta_2^3)^{-3} \de \beta_2^3$ occurring in \eqref{naturab} 
will be corrected to
$\de \beta_2^3 / \beta_2^3$ --- the desired pole --- once we include an extra factor
$(\beta_2^3)^{2}$ which comes from the free-fermion correlator $\omega$.
We now turn to the analysis of this factor.


\subsubsection{Factors from the fermion correlator}

The integrand \eqref{dtwoc} contains the factor
\begin{equation} \label{free-fermions-copy}
\omega(\sigma_1, \dots, \sigma_n) =
\prod_{i=1}^n \frac{\de\sigma_i}{\sigma_i - \sigma_{i+1}},\qquad
\sigma_{n+1}\equiv \sigma_1.
\end{equation}
We would like to investigate how this form behaves on
conics that are degenerating into a pair of lines ({\it i.e.}\ near $\Mint$.)
The result will be that along $\Mint$, $\omega$ factorizes into a product
of two copies of $\omega$ defined on the two lines separately (with an extra
marked $\sigma$ on each line at the point of intersection), while transverse
to $\Mint$, $\omega$ vanishes like $(\beta_2^3)^{2}$.
 
As the curve degenerates to
a pair of lines, some of the $n$ insertions approach one line and some approach the
other.
We consider the case where
\eqn{firstgroup}{\sigma_1, \dots,\sigma_m}
approach one asymptote while the remaining
$(n-m)$ insertions 
\eqn{secondgroup}{\sigma_{m+1}, \dots, \sigma_n}
approach the other.  This is not the most general choice, since the $\sigma_i$ come
with a fixed cyclic ordering which is built into \eqref{free-fermions-copy}; our
choice is characterized by the fact that as we run through the cyclic ordering we jump from the first
line to the second and back only once.
We will comment on other possibilities at the end.

With the $GL(2,\C)$ gauge-fixing we chose above, 
as we approach some point of $\Mint$, the coordinates $\sigma_i$ do not remain finite;
one of the lines is $\sigma \to 0$ while the other line is $\sigma \to \infty$.
So we need to rescale the $\sigma_i$ to get new coordinates $\hat \sigma_i$ on $\Mint$
which label the positions of the marked points;
we define $\hat \sigma_i$ so that 
$Z^\IA$ defined in \eqref{asymplines} remains constant
as $\hat\sigma_i$ is kept fixed and $\beta_0^3,\beta_2^3 \to 0$. 
The correct redefinition is
\eqn{newsigma}{\sigma_i
=
\left\{
\begin{array}{lcl}
(\beta_2^3)^{-1} \hat\sigma_i &\mbox{~for~}& i\in\{1,2,\dots m\}\\
\beta_0^3 (\hat\sigma'_i)^{-1} &\mbox{~for~}& i\in\{m+1, 
m+2,\dots n\}\\
\end{array}
\right\}.
}
(We use two different symbols $\hat\sigma_i$ and $\hat\sigma'_i$
to distinguish the coordinates on the two different lines.)
Rewriting $\omega$ from \eqref{free-fermions-copy} in terms of
$\hat\sigma_i$ and $\hat\sigma'_i$, we obtain
\eqn{newsigmaomega}{
\omega(\hat\sigma_1, \dots \hat \sigma'_n)
=
\beta_0^3\beta_2^3
\left(\prod_{i=1}^{m-1}
\frac{\de \hat\sigma_i}{\hat\sigma_i-\hat\sigma_{i+1}}\right)
\frac{\de \hat\sigma_m}{\hat\sigma_1\hat\sigma_m}
\left(\prod_{i=m+1}^{n-1}
\frac{\de \hat\sigma'_i}{\hat\sigma'_i-\hat\sigma'_{i+1}}
\right)
\frac{\de\hat\sigma'_n}{\hat\sigma'_{m+1}\hat\sigma'_n}+ \dots
}
where the intersection was defined to be at $\hat\sigma=\hat\sigma'=0$.
The dots in \eqref{newsigmaomega}
indicate terms suppressed by powers of $\beta_0^3\beta_2^3$.

Most of the powers of $\beta^3_0$ and $\beta^3_2$ have canceled,
but there is an extra factor of $\beta_0^3\beta_2^3$,
which equals $(\beta_2^3)^2$ because of our gauge
choice \eqref{gaugechoicee}.  Also, we obtained the expected free fermion contractions, 
including the $2+2$ 
contractions 
involving the intersection of the
two lines at
$\hat\sigma=\hat\sigma'=0$.

Note that $\beta_0^3$ and $\beta_2^3$ always appeared in the combination
\eqn{ccombi}{C=\beta_0^3\beta_2^3}
that is invariant under 
\eqref{unfixed}.
This is the same $C$ that we used 
in Figure \ref{branches};
in fact, one can rewrite our curve in the form
\eqn{xyc}{xy=C}
where $x$,$y$ are coordinates on a plane in $\CPP$.
The limit $C\to 0$ describes the singular conics.
Note that it is $C$ rather than $\beta_2^3$ that is a good coordinate ---
this is because a simultaneous sign flip on $\beta_0^3$ and $\beta_2^3$ 
is the gauge transformation 
\eqref{unfixed} with $\lambda = i$,
which preserves our gauge choices \eqref{gaugechoicee}.

Finally, it is easy to check that if we choose a different distribution of
the marked points, the result comes out suppressed by additional powers of $C$.
We are only interested in the leading terms, which are linear in $C$ and will give 
the coefficient of \,$\de C / C$.

\subsection{Finishing the proof in degree 2 case} \label{finishing}

Now we can collect the results from the previous two 
subsections.  
The powers of $\beta_2^3$ from 
\eqref{naturab} and \eqref{newsigmaomega} combine to give
$\smallint \de \beta_2^3 / \beta_2^3$, which is proportional to $\smallint
\de C / C$.  So as advertised, the integral \eqref{dtwoc} localizes
after contour integration to an integral over $\Mint$.  The symmetries of 
$\CPP$ determine the measure for the moduli of the two lines in $\Mint$, which therefore agrees
with the measure $\muint$ in \eqref{dtwoi} up to an overall constant; as for
the integral over the marked points, comparing \eqref{newsigmaomega} and
\eqref{dtwoi} we see that these measures are also identical.  This completes
the argument for equivalence in the $d=2$ case.

Incidentally, one can also compare the measures on $\Mint$ directly, without
recourse to a symmetry argument.  
We have already computed
the measure which arises from the connected prescription, in \eqref{naturab},
so the job is to compute the measure $\muint$ which arises from the disconnected
prescription.  This computation is given (in greater generality) in section \ref{finishing-higher}.


\section{Higher degree \label{higherdegree}}

Now let us consider the connected prescription for general degree $d$.
We will see that the fully disconnected description and the fully connected
prescription are not only equivalent, they are
just two extreme cases of a more general class of
rules to calculate the amplitude.
We will find $d$ \ti{a priori} different expressions for
the scattering amplitude with $d\!+\!1$ negative-helicity
gluons,
\eqn{ways}{{\cal A}_{[K]},\qquad K=0,1,2,\dots , d-1,}
where $K\!+\!1$ denotes the total number of curves involved in the 
prescription.\footnote{Later we will see that $K$ also represents the codimension in moduli space on which
the prescription is localized, or equivalently the number of internal propagators
which appear in the prescription.}

The organization of this section is as follows:

\begin{itemize}

\item subsection \ref{plethora} outlines the argument that
the completely connected and completely disconnected prescriptions
agree;

\item subsection \ref{csw-generalized} discusses the intermediate prescriptions
with arbitrary $K$ and their diagrammatic interpretation;

\item subsection \ref{higher-residue} generalizes the residue
calculation of subsection \ref{lower-residue} to the case of a degree $d$
curve splitting into two curves of degrees $d_1$ and $d_2$;

\item subsection \ref{finishing-higher} shows that the residues occurring
for any degeneration are actually independent of the chosen prescription, 
completing the argument.

\end{itemize}

\subsection{The proof in higher degree case\label{plethora}}

Rather than showing directly that the connected prescription arising from a single
connected degree $d$ curve is equivalent to the disconnected
prescription involving $d$ lines, we will first show that it is
equivalent to a computation 
involving two disconnected components of degrees $d_1$, $d_2$, such that 
\begin{equation} \label{dd}
d_1 + d_2 = d.
\end{equation}
The proof is a 
generalization of the computation we did in section \ref{lower-residue}:  
namely, in subsection \ref{higher-residue} we will find a pole 
on each boundary divisor $\Mintgamma$, corresponding to a degeneration into
intersecting curves,
\begin{equation}
\Sigma_d \longrightarrow \Sigma_{d_1} \cup \Sigma_{d_2},
\quad \quad \quad d_1 + d_2 = d,
\label{dddcurves}
\end{equation}
with a particular distribution of the marked points.  

\begin{figure}[t]
\begin{center}
\epsfig{file=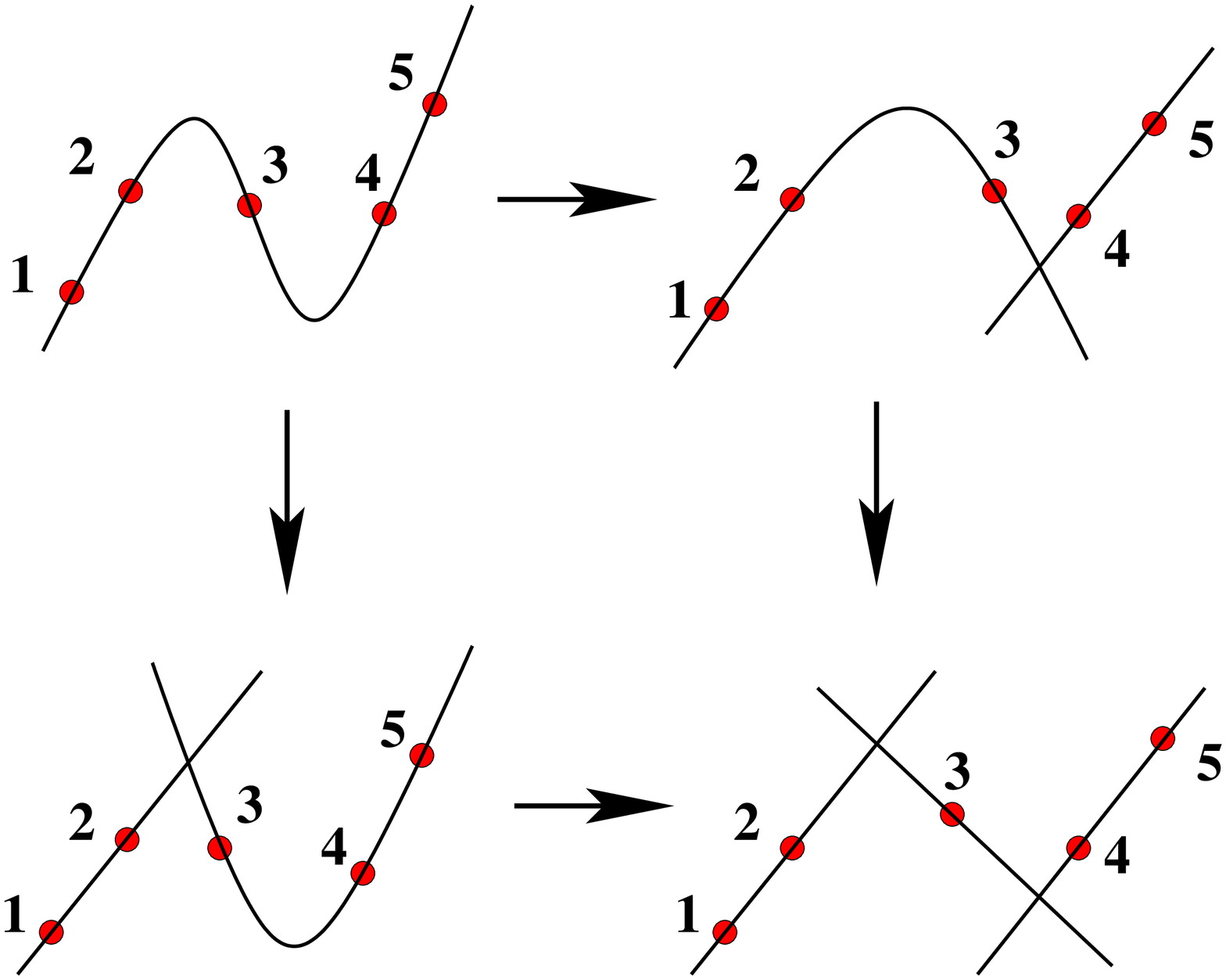,width=120mm}
\end{center}
\caption{A degeneration of a degree 3 curve into three intersecting
lines can be viewed as a two-step process.
The moduli space of degree 3 maps with 5 marked points, $\Mstab{5}{3}$,
contains divisors, $\M^{\Lambda_1}_{{\rm int}}$ and $\M^{\Lambda_2}_{{\rm int}}$,
associated with degenerations into a degree 2 curve and a line,
shown at the intermediate stages. The moduli space $\Mintgamma$
of three intersecting lines (shown in the lower right corner) can be
identified with the intersection $\M^{\Lambda_1}_{{\rm int}} \cap \M^{\Lambda_2}_{{\rm int}}$.}
\label{d3curves}
\end{figure}

Next we want to show iteratively that
this integral over curves with $2$ irreducible components is equivalent to one over curves with $3$ components, and so on until eventually we reach $d$ components (all of which must
have degree $1$.)  The idea which makes this iteration possible is the following:  
consider some locus
$\Mintgamma$, corresponding to a particular degeneration of $\Sigma$ into
$K\!+\!1$ components, with a particular distribution of the marked points.
This locus
can be obtained as an intersection of $K$ boundary divisors, $\M^{\Lambda_j}_{{\rm int}}$,
each of which is associated with a degeneration of $\Sigma_d$
into two irreducible components,\footnote{We use $\Gamma$ to denote a general 
degeneration into $K\!+\!1$ components, and $\Lambda$ to denote a degeneration into just two components.}
\begin{equation}
\Mintgamma = \M^{\Lambda_1}_{{\rm int}} \cap \cdots \cap \M^{\Lambda_{K}}_{{\rm int}}.
\label{mintascap}
\end{equation}
An example is shown in Figure \ref{d3curves}.  
In this sense, the problem 
of studying a general degeneration boils down to
understanding the basic process \eqref{dddcurves}.

So let's start with the integral over 
$K$-component curves and try to prove it agrees
with an integral over $(K+1)$-component curves.  
In the $K$-component case we
have to integrate over various loci $\Mintgamma$ as in
\eqref{mintascap}.
Since the various divisors $\Mintlambda$ meet transversally \cite{MR98m:14025},
in integrating over each such $\Mintgamma$ we will encounter poles
wherever $\Mintgamma$ intersects another divisor $\Mintlambda$.\footnote{One
way to understand this is to note that if we start with
the full $\Mstab{n}{d}$ and look near such an intersection of $K$ divisors,
the integrand looks like
\begin{equation}
\frac{\de C_1}{C_1} \wedge \cdots \wedge \frac{\de C_{K}}{C_{K}}
\wedge ({{\rm regular}}).
\end{equation}
We have already contour-integrated 
over $C_1, \dots, C_{K-1}$ and thus restricted to $C_1 = \cdots = C_{K-1} = 0$, \ti{i.e.}\ 
to $\Mintgamma$; after doing this we get simply $\de C_{K}/C_{K}$, with a pole at $C_K = 0$, 
\ti{i.e.}\ at $\Mintgamma \cap \Mintlambda$.}
We choose our contour so that
it picks up the residues at these poles.
In this way we reduce the integral over $\Mintgamma$ to the sum
of integrals over all intersections $\Mintgamma \cap \Mintlambda$.
Then we have to sum over all $\Gamma$ describing $K$-component degenerations.
What is the result of all this summation? From the perspective
of the $(K\!+\!1)$-component degenerations --- which we label by $\Gamma'$ ---
the answer is clear: given some 
\begin{equation}
\Mintgammaprime = \M^{\Lambda_1}_{{\rm int}} \cap \ldots \cap \M^{\Lambda_{K}}_{{\rm int}},
\label{mintascap2}
\end{equation}
there are $K$ ways to make it by intersecting
some $\Mintgamma$ with some $\M^{\Lambda_i}_{{\rm int}}$.
Therefore we get a sum over all $(K\!+\!1)$-component
degenerations, with an \ti{overall} multiplicative factor $K$.

Finally, after repeating this process $d-1$ times, we arrive at an 
integral over the moduli space of connected trees consisting of $d$ lines,
with all possible shapes for the tree and all allowed distributions of
marked points.  But the arguments of \cite{Cachazo:2004kj} show that the disconnected prescription also reduces to such an integral, by a similar process of localization
to poles.  Furthermore, in section \ref{higher-residue}
we will see that the residues in these two computations agree;
this will complete the proof.

\subsection{Intermediate prescriptions} \label{csw-generalized}

In subsection \ref{plethora} we encountered $d-1$ different moduli spaces $\MintK$ of singular curves, characterized by the number $K+1$ of components, 
which interpolated between the nonsingular degree $d$ curve ($K=0$) and the tree of degree $1$
curves ($K=d-1$).  Furthermore we obtained integrals over each $\MintK$ by starting with the connected
prescription ($K=0$) and successively localizing to poles.  As a result of this localization
all these integrals are equal; now we want to argue that the intermediate cases $K=1, \dots, 
d-2$ can be naturally interpreted as coming from ``intermediate prescriptions,'' involving
integrals over the moduli of $K+1$ disconnected curves with $K$ propagators connecting them.  
We defined these prescriptions at the end of section \ref{review-disconnected}.

The argument is a generalization of the ``heuristic'' derivation of
the computational rules for the disconnected prescription, given 
in \cite{Cachazo:2004kj}.
Namely, starting from the intermediate prescription,
note that the propagator $D(\cdot, \cdot)$ by definition satisfies
\begin{equation} \label{prop-def}
\bar{\partial} D = \Delta.
\end{equation}
Here $\Delta$ is a $(0,3)$-form on $(\CPP)^2$
which is concentrated on the diagonal $\CPP$:
in inhomogeneous coordinates with $Z^4 = Z'^4 = 1$ it may be written
\begin{equation}
\Delta = \deltabar(Z^1 - Z'^1)\,\deltabar(Z^2 - Z'^2)\,
\deltabar(Z^3 - Z'^3)\,\delta^{4}(\psi - \psi'),
\qquad
\deltabar(f) := \delta^2(f) \de\bar{f}.
\end{equation}
The equation \eqref{prop-def} means that $D(\cdot, \cdot)$
is meromorphic with a pole along the diagonal.  
The integral over $\MintK$ in the disconnected prescription contains $K$
propagators \eqref{internal}; these factors therefore
have poles when $Q_i(\sigma) = Q_{i'}(\sigma')$.  

As in \cite{Cachazo:2004kj}, we assume that $K$ of the integrals over
moduli of the disconnected curves are evaluated on contours
which encircle these poles, in a suitable sense.
Using \eqref{prop-def}, performing these contour
integrals is equivalent to filling in the contour and 
replacing $D$ by $\Delta$.
This localizes the integral to the sublocus of moduli
space where all propagators have shrunk to zero length,
which is exactly $\MintK$.

So finally we have $d$ different prescriptions, involving summing 
over configurations with $1$ curve (connected case), $2$, $3$, \dots, $d$ curves (maximally disconnected case); and we have argued that each of these prescriptions is equivalent, up to an overall rescaling.  In this sense any of them can be used to calculate the Yang-Mills
amplitudes.

Of course, another possibility is that the correct amplitudes are obtained by summing
different contributions from various sorts of diagrams with various numbers of curves.  We have argued that all such contributions are proportional to one 
another, so such a modified rule would only change the overall prefactor.
Although we will not try to make the final verdict in this paper, we believe that a more
detailed analysis of the prescriptions (including the coefficients) should be able
to resolve this uncertainty.

\subsubsection{Diagrammatic interpretation and an example}

Now let us discuss the diagrammatic interpretation of the
intermediate prescriptions.
We have seen that the $K$-th intermediate prescription is naturally localized
on $\MintK$, which is a union of various $\Mintgamma$.  Here $\Gamma$
describes the decomposition of the curve $\Sigma_d$ into $K+1$ components
and the distribution of marked points along these components.  
Equivalently, we could say that
$\Gamma$ describes a slight generalization of an MHV tree diagram:
namely, it is a tree diagram with $K+1$ vertices, where each vertex now carries
an internal index $d_i$, subject to the rule that $\sum d_i = d$.  The MHV
diagrams are the case where all $d_i = 1$.

\begin{figure}
\begin{center}
\epsfig{file=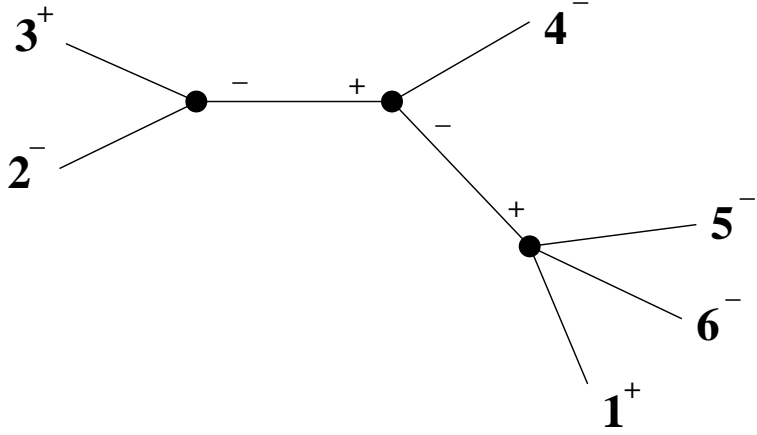,width=100mm}
\end{center}
\caption{An MHV tree diagram contributing to $\A{+-+---}$.}
\label{twistor3}
\end{figure}

It would be very useful if we could give a compact formula for the contribution
of a general vertex with arbitrary $d_i$, analogous to the off-shell
continuation of the MHV amplitude given in \cite{Cachazo:2004kj}.
At the moment we do not possess such a formula, so we can only define the diagram
$\Gamma$ to be the integral over $\Mintgamma$ which we considered above.  In this
language, our localization argument relating different prescriptions 
becomes the statement that the contribution from a diagram $\Gamma$ agrees with
the sum over all $\Gamma'$ obtained by ``splitting a vertex'' in $\Gamma$.  In
other words, $\Gamma'$ should be obtained by replacing a 
vertex with index $d$ by a pair of vertices with indices
$d_1$, $d_2$, such that $d_1 + d_2 = d$, with a propagator connecting them.  This is
the diagrammatic analog of a degree $d$ curve which degenerates into two curves
with degrees $d_1$, $d_2$.

We can also repeat the combinatorics from
subsection \ref{plethora} in this language.  Start with a diagram with $K\!+\!1$
vertices.  This diagram contains $K$ propagators.
Therefore there are $K$ ways to shrink a single propagator
and obtain a ``parent'' diagram with $K$ vertices.
Because a diagram with $K\!+\!1$ vertices has $K$ parents, 
the sum over the daughters 
with $K\!+\!1$ vertices equals $K$ times the sum over
the parents with $K$ vertices. 

\begin{figure}[t]
\begin{center}
\epsfig{file=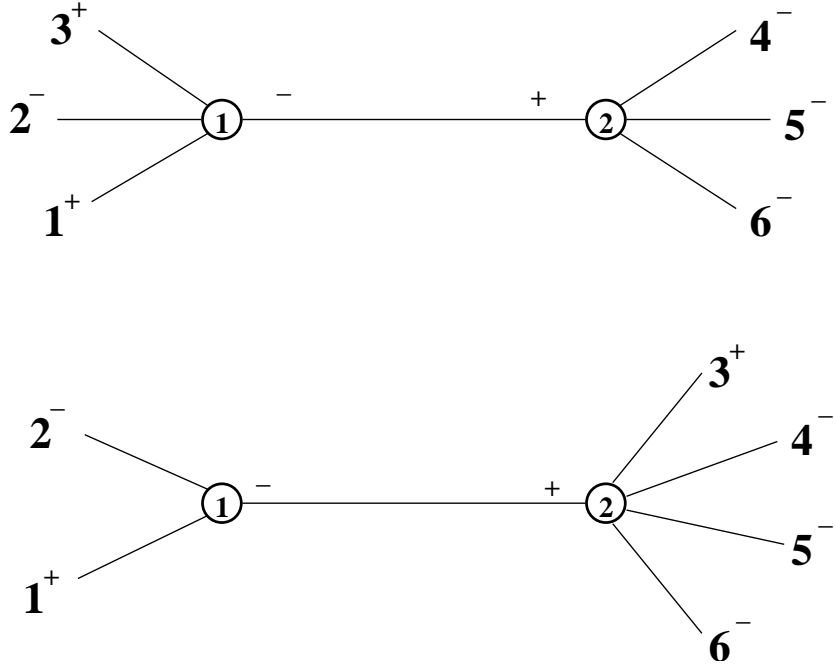,width=110mm}
\end{center}
\caption{Two types of tree diagram with one MHV and one
non-MHV (degree 2) vertex that contribute to the $\A{+-+---}$ amplitude.
In total, there are six diagrams of each kind.
The number attached to each vertex represents the degree
of the corresponding curve in twistor space.}
\label{twistor2}
\end{figure}

To illustrate how all this works when external wavefunctions of fixed
helicity are included, let us consider a 6-gluon amplitude $\A{+-+---}$.
If we were to use the connected prescription, we would have
to integrate over the moduli space $\Mstab{6}{3}$ of degree 3 curves.
On the other hand, in the disconnected prescription one has
to consider three degree 1 curves, which can be interpreted
as MHV vertices in Yang-Mills theory \cite{Cachazo:2004kj}.
Therefore, in this case one has to sum over all tree graphs with
three MHV vertices connected by Yang-Mills propagators --- see Figure \ref{twistor3}.
In total, there are 19 such graphs contributing to $\A{+-+---}$.

Now let us consider the intermediate prescription with $K=1$. 
This prescription leads
to a sum over tree graphs with two vertices, one MHV
and one non-MHV (the non-MHV vertex involves three insertions of negative helicity).
Examples of such graphs with non-MHV vertices are shown in Figure \ref{twistor2}.
There are 12 such diagrams which contribute to $\A{+-+---}$.
Since each non-MHV vertex itself can be represented
as a sum over tree diagrams with two MHV vertices, we should be
able to reproduce the disconnected prescription if we split
all non-MHV vertices into MHV ones.
More precisely, in this decomposition we should encounter each
MHV diagram twice (since in the disconnected prescription $K=2$).
Indeed, it is straightforward to check that the 12 non-MHV diagrams
lead to 38 MHV graphs, in agreement with the general rule.

\subsection{Computing the residue in higher degree case\label{higher-residue}}

Returning from our digression to discuss the intermediate prescriptions, 
in this section we show that the integral \eqref{formal-yma} over the
moduli space $\Mstab{n}{d}$ which arises in the connected prescription
has a pole along the codimension 1 divisor $\Mintone$ describing
curves that are degenerated into 2 components.  We further verify
that the residue is the same as that which arises after localization of the $K=1$ 
prescription on $\Mintone$, thus establishing the equivalence between connected
and $K=1$ prescriptions.

We want to study a degeneration in which 
the curve $\Sigma_d$ degenerates into a pair of intersecting curves,
$\Sigma_{d_1}$ and $\Sigma_{d_2}$, of degree $d_1$ and $d_2$,
as in \eqref{dddcurves}.
Using the projective symmetry to divide by $\sigma^{d_1}$,
we can write the degree $d$ map \eqref{polyno} as
\eqn{polynood}{
Z^\IA(\sigma) = \sum_{k= -d_1}^{d_2} \beta^\IA_{d_1+k} \sigma^k.
}
We fix the $GL(2,\C)$ symmetry similarly to 
the degree 2 case,
namely by conditions based on \eqref{gaugechoice} and \eqref{gaugechoicee}:
\eqn{fixing-d}{
(\beta^4_{d_1-1},\beta^4_{d_1},\beta^4_{d_1+1}) = (0,1,0),\qquad
\frac{\beta^3_{d_1-1}}{\beta^3_{d_1+1}} = 1,
}
and define the deformation parameter $C:= \beta_{d_1-1}^3 \beta_{d_1+1}^3$.
As in degree $2$, the singular limit will be $C \to 0$, or equivalently
$\beta_{d_1+1}^3 \to 0$, and the question is
how the other coefficients should scale in this limit.

\begin{figure}
\begin{center}
\epsfig{file=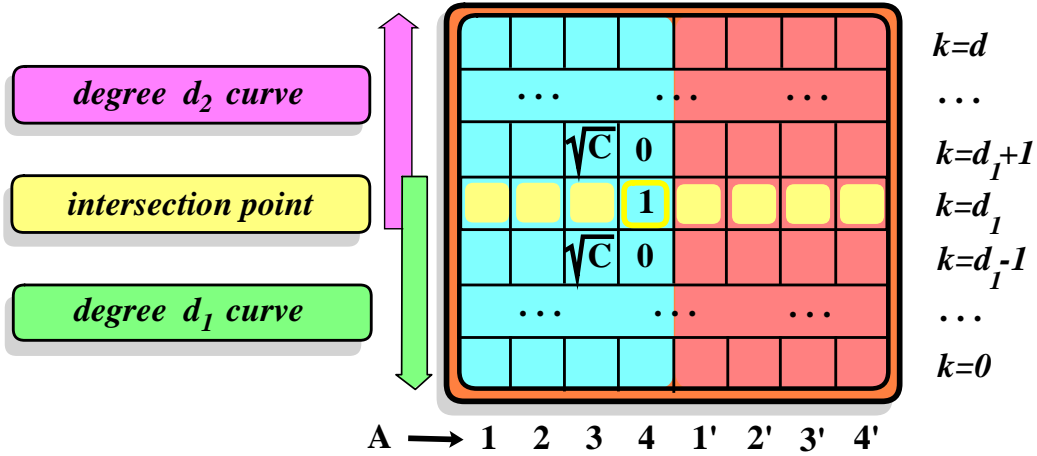,width=140mm}
\end{center}
\caption{The organization of the coefficients $\beta^\IA_k$
for a degree $d$ curve degenerating into curves of degrees $d_1$ and $d_2$.
The symmetry $GL(2,\C)$ is fixed by setting three bosonic coefficients to the values
$(0,1,0)$ and two others to $\sqrt{C}$; 
this $C$ is the deformation parameter, which approaches zero in the degeneration
limit.
}
\label{table-beta}
\end{figure}

The correct scaling is as follows:  we take $\beta_{d_1-1}^3 = \beta_{d_1+1}^3 \to 0$ while holding
finite the quantities
\eqn{coef-done-dtwo}{
\alpha_k^\IA := \frac{\beta^\IA_{d_1-k}}{(\beta^3_{d_1-1})^k}, \quad 0\leq k \leq d_1; \qquad
\alpha'^\IA_k := \frac{\beta^\IA_{d_1+k}}{(\beta^3_{d_1+1})^k}, \quad 0\leq k \leq d_2.}
In that limit we obtain two curves,
\begin{eqnarray} \label{curves-new-coords}
\Sigma_{d_1}~: && Z^\IA(\hat\sigma) = \sum_{~k= 0~}^{d_1} 
\alpha_k^\IA \hat\sigma^{k}, \nonumber \\
\Sigma_{d_2}~: && Z^\IA(\hat\sigma') = \sum_{~k= 0~}^{d_2} 
\alpha'^\IA_k \hat\sigma^{\prime k}.\label{two-curves-b}
\end{eqnarray}
Namely, we obtain the points $Z^\IA(\hat\sigma)$ on $\Sigma_{d_1}$ by holding fixed
$\hat\sigma = \sigma / \beta^3_{d_1-1}$ in the limit, 
and we obtain the points $Z^\IA(\hat\sigma')$ on $\Sigma_{d_2}$ by holding fixed
$\hat\sigma' = \sigma \beta^3_{d_1+1}$ in the same limit.
See Figure \ref{table-beta}.

Therefore the parameters $\alpha_k^\IA, \alpha'^\IA_k$ give coordinates on $\Mintone$,
specifying the moduli of the degenerated curve.
(Note
that $\alpha_0^\IA = \alpha'^\IA_0$; these shared coordinates 
specify the intersection
point of $\Sigma_{d_1}$ and $\Sigma_{d_2}$.)

Now we want to study how our integral \eqref{formal-yma} behaves near $\Mintone$.
As in section \ref{lower-residue}, we have to compute the Jacobian $J$ from the
gauge-fixing of $GL(2,\C)$.  The matrix of variations generalizing \eqref{fixingmatrix}
is
\eqn{fixingmatrix-2}{
\delta
\left(
\begin{array}{c}
\beta^4_{d_1-1}\\
\beta^4_{d_1}\\
\beta^4_{d_1+1}\\
\beta^3_{d_1-1} / \beta^3_{d_1+1}
\end{array}
\right)
=
\left(
\begin{array}{ccrr}
d_2&\!\!\!\!(d_1+2)\beta^4_{d_1+2}&0&0\\
\!\!\!\!(d_2+2)\beta^4_{d_1-2}\!\!&d_1 &0&0\\
0&0&d_1\!&d_2\! \\
*&*&2&-2
\end{array}
\right)
\left(
\begin{array}{c}
b\\
c\\
a\\
d
\end{array}
\right).
}
In the singular limit, the $\beta^4_{d_1 \pm 2}$ terms in \eqref{fixingmatrix-2} 
vanish, and we get
\begin{equation} \label{jac-1}
J \to -2 d_1 d_2 d.
\end{equation}
The gauge-fixed integral includes the factor $J / 2d$; the $2d$ comes from
an unfixed subgroup of $GL(2,\C)$, analogous to \eqref{unfixed-discrete-matrix}, 
which is $\Z_2 \times \Z_d$ if both
$d_1$ and $d_2$ are even and $\Z_{2d}$ otherwise.
Next we have to rewrite the integrand in terms of the new coordinates \eqref{coef-done-dtwo}.
One might be worried that switching to these coordinates
will generate extra powers of $C$ beyond what we had in the degree 2 case,
spoiling the conclusion that there is a pole along $\Mintone$.
But this does not occur; if we increase $d_1$ by 
$1$, for example, the integrand just acquires an extra integral over $4|4$ variables:
\eqn{measure-new}{ \mu \to \mu \wedge \prod_\IA \de \beta^\IA_{0} =
\mu \wedge
\prod_\IA \de \alpha^\IA_{d_1}
}
The powers of $\beta^3_{d_1+1}$ simply cancel between the 4 bosons and 4 fermions!
Unlike the coefficients $\beta^\IA_{d_1}$ and
$\beta^\IA_{d_1\pm 1}$, among which $5$ special bosonic components
have been used to gauge-fix the $GL(2,\C)$ symmetry or to describe the parameter $C$,
the additional moduli $\beta^\IA_{d_1\pm k}$ with $k>1$ come in full ``supermultiplets''
containing 4 bosons and 4 fermions.  Therefore no new powers of $C$ are generated in
rescaling $\beta$'s to $\alpha$'s,
so the measure for the moduli of the degenerating curve still 
behaves as $\de C / C^2$ near $C=0$.  Similarly, the free
fermion correlator $\omega$ factorizes,
\begin{equation}
\omega(\sigma) \to C\,\omega_1(\hat\sigma) \wedge \omega_2(\hat\sigma'), 
\end{equation}
just as in degree $2$.

So we have a pole along $\Mintone$, as in the degree $2$ case, and after integrating
around this pole the fully gauge-fixed measure
for the moduli of the degenerate curve is
\begin{equation} \label{measure-degenerate}
-d_1 d_2 \prod_{\IA} \left( {\prodprime{k=0}{d_1}}\! \de \alpha_k^\IA\,{\prodprime{k=1}{d_2}}\! \de \alpha'^\IA_k \right).
\end{equation}
Here the symbol $\Pi'$ indicates that we omit the $5$ factors
\begin{equation} \label{fixed-components}
\de \alpha^4_1,\, \de \alpha'^4_1,\, \de \alpha^4_0,\, \de \alpha^3_1,\, \de \alpha'^3_1;
\end{equation}
there are no such $\alpha$'s among the coordinates on $\Mintone$, because their corresponding $\beta$'s
were used up in the gauge-fixing and in the transverse coordinate $C$, as shown in Figure \ref{table-beta}.

\subsection{Finishing the proof in higher degree case} \label{finishing-higher}

Finally we
have to check that the measure \eqref{measure-degenerate} agrees with the one coming from localization
of the $K=1$ prescription.  From section \ref{csw-generalized} we know that the latter measure
is obtained as follows:  start with two curves of degree $d_1$, $d_2$,
\begin{align} 
Q^\IA(\sigma) &= \sum_{k=0}^{d_1} \alpha^\IA_{k} \sigma^k, \nonumber \\
Q'^\IA(\sigma') &= \sum_{k=0}^{d_2} \alpha'^\IA_{k} \sigma'^k. \label{dcurves}
\end{align} 
(The notation $\alpha^\IA_{k}$, $\alpha'^\IA_k$ we use here
agrees with the notation we
used above for the moduli of the curves obtained by a degeneration; compare \eqref{dcurves} with \eqref{curves-new-coords}, \eqref{coef-done-dtwo}.  The only difference is that here we do not necessarily have $\alpha^\IA_0 = \alpha'^\IA_0$.)
Then we have the standard measure \eqref{natural} on the two curves separately, which before gauge-fixing is
\begin{equation} \label{standard}
\mu_{d_1} \wedge \mu_{d_2} = \prod_{\IA} \left( \prod_{k=0}^{d_1} \de \alpha^\IA_k \prod_{k=0}^{d_2} \de \alpha'^\IA_k \right).
\end{equation}
As explained in section \ref{csw-generalized}, the requirement that the two curves actually intersect is enforced by a delta function which is coupled to one marked point on each curve,
\begin{equation} \label{deltafunc}
\Delta(Q(\sigma), Q'(\sigma')).
\end{equation}
To compare the measures (including this delta function) we have to gauge-fix
the $GL(2,\C)^2$ symmetry acting on the coefficients $\alpha^\IA_k, \alpha'^\IA_k$.
There are many ways to do this;
we choose a way that is as similar as possible to the gauge-fixing we used for the degenerating
degree $d$ curve, so that the unfixed moduli will match directly.
Namely, we take
\begin{align} \label{gauge-conditions1}
\alpha^i_0 &= \alpha'^i_0 \quad \textrm{for}\ i \in \{2,3\}, \\
\alpha^4_0 &= \alpha'^4_0 = 1, \label{gauge-conditions2} \\
\alpha^4_1 &= \alpha'^4_1 = 0, \\
\alpha^3_1 &= \alpha'^3_1 = 1.
\end{align}
The matrix of variations from this gauge-fixing is similar to \eqref{fixingmatrix-2},
but since it is an $8 \times 8$ matrix we just write the answer here:
\begin{equation} \label{jac-2}
J = (d_1 d_2)^2 (\alpha^2_1 - \alpha'^2_1).
\end{equation}
The gauge-fixing factor is $J / {d_1 d_2}$, because of the subgroup $\Z_{d_1} \times \Z_{d_2} \subset
GL(2,\C) \times GL(2,\C)$,
roots of unity acting on each curve separately; since this subgroup
acts trivially it is unfixed by our gauge condition. 
Next we must include the integral over the delta function \eqref{deltafunc}, which we write as
\begin{equation}
\int \de \sigma\ \de \sigma'\ \delta^{(3|4)} \left( \frac{Q^\IA(\sigma)}{Q^4(\sigma)} - \frac{Q'^\IA(\sigma')}{Q'^4(\sigma')} \right).
\end{equation}
With our gauge choice, it is easy to study the behavior of this delta function in the vicinity
of $\sigma = \sigma' = 0$.\footnote{Although our gauge choice was rigged so that studying 
$\sigma = \sigma' = 0$ would recover the desired moduli space of intersecting curves, 
it is not clear \ti{a priori} 
from our arguments why one should consider only this region; this is related to the issue of the
exact contour choice in the intermediate prescription, which we will not settle here.  We are also integrating over the delta function as if it were real instead of holomorphic;
similar manipulations were used in \cite{Witten:2004cp}.}  
One uses the $Z^2$ and $Z^3$ components of the delta function to set $\sigma = \sigma' = 0$, 
obtaining
\begin{equation} \label{deltas}
\frac{1}{(\alpha^2_1 - \alpha'^2_1)}\ \delta(\alpha^1_0 - \alpha'^1_0) \prod_{\IA=1'}^{4'} \delta(\alpha^\IA_0 - \alpha'^\IA_0).
\end{equation}
Note that the $1|4$ delta functions in \eqref{deltas}, combined with the gauge conditions \eqref{gauge-conditions1}, \eqref{gauge-conditions2}, 
are enough to set all $\alpha'^\IA_0 = \alpha^\IA_0$.  This was
the main motivation for this gauge-fixing; the point $\alpha^\IA_0$ represents the intersection 
of the two curves, and the remaining moduli are precisely the ones we had for the degenerating degree
$d$ curve in \eqref{measure-degenerate}.  Therefore we easily see that the measures agree, including
the prefactor $d_1 d_2$.
(Although we have not been careful about overall constant factors,
the absence of a relative factor here is important --- it corresponds to the absence of prefactors
weighing different diagrams in the intermediate prescriptions.)

This completes the argument for the equivalence between the connected and $K=1$ prescriptions.
It also sets up the iteration we described in section \ref{plethora} to prove
the equivalence of all prescriptions, by successive localization to poles in higher
and higher codimension, corresponding to more and more degenerate curves.

One detail remains:
we have to check that the residues we obtain are always independent of which
prescription we started with.  
In other words, we have to prove that
the measure for the integral over $K+1$-component trees obtained by some degeneration process
always agrees with the measure coming from the disconnected prescription.  
As we know from section \ref{csw-generalized}, the latter measure
can be written as a product of measures for the individual curves, with delta-functions
that guarantee the curves intersect.  
We just proved the agreement for $K=1$.  For general $K$ we can work 
inductively; given a $K+1$-component tree
on which some curve is further degenerating, just focus on the measure for that curve, and 
note that the delta-functions from the other curves are well behaved on moduli space
near the degeneration we are studying.  In this sense the degenerating curve can
be isolated from the rest of the tree.  The computation done above in the $K=1$ case then 
shows that the measure after this degeneration agrees with that from the disconnected 
prescription.  This then completes the argument for the equivalence of all prescriptions.

\section{Conclusions and open questions}

We have argued for the equivalence of the connected and disconnected
twistorial formulae for the tree level scattering amplitudes of $\N=4$ super Yang-Mills,
provided that the contours are appropriately chosen.
Using this equivalence we can now exploit the complementary virtues of the two
prescriptions simultaneously.  As we remarked in the introduction, the connected prescription
minimizes the number of diagrams one has to sum, namely, there is only one; the amplitude 
is expressed as a single integral, which was the starting point for several theoretical developments
\cite{Berkovits:2004hg,Berkovits:2004tx,Witten:2004cp}.  The disconnected
prescription involves more diagrams, but still a manageable number for some interesting
amplitudes, and the contribution from each diagram can be immediately written down.  

To conclude, we summarize some of the many remaining open problems in this area:

\begin{itemize}

\item {\bf Contours I.}
Is there a rigorous justification of the choice of contours in all these calculations?
In our argument for the equivalence between connected and disconnected
prescriptions we identified specific poles in the integral
over moduli, and we roughly wanted a contour which encircles all of these poles.
We believe it should be possible to show by a deformation argument that 
our choice of contour is equivalent to the one used in \cite{Roiban:2004yf},
thus completing the proof of equivalence,
but this seems to be nontrivial; the 
computations in \cite{Roiban:2004yf} depend on a particular method of evaluating the integral
in the connected prescription by saturating delta-functions, and it is difficult to see which contour 
it corresponds to.

\item {\bf Contours II.} Once the residues are isolated in both prescriptions,
we must still integrate over the degeneration locus 
$\Mint$, which requires yet another choice of
contour; for example, the integration over $t$ from $0$ to $\infty$ in section
6 of \cite{Cachazo:2004kj} should have some \ti{a priori} justification. 
This paper has not addressed this question.
Our argument for the equivalence requires that the contours on $\Mint$ are chosen
to be equivalent in all prescriptions.

\item {\bf Explicit external wavefunctions.}
Our derivation was rather formal. It did not depend on the particular form of the wavefunctions.
Of course, it would be interesting to verify the picture by calculating the amplitudes 
involving particles with well-defined momenta {\it i.e.}\ $(\lambda,\tilde\lambda,\psi)$
using our generalized prescriptions.  Unlike the MHV vertices in \cite{Cachazo:2004kj},
one might expect that the $d>1$ vertices will be ratios of polynomials involving both $\lambda$ and $\tilde\lambda$.  (Of course, it is also possible that one will not 
obtain any compact formula for the $d>1$ vertices in this way.)

\item {\bf Derivation from the B-model.}
Both connected and disconnected contributions seem to arise in the topological
B-model of \cite{Witten:2003nn} as long as we use not only the degree $d$ 
D1-instantons but also the propagators (and vertices) of the holomorphic Chern-Simons theory.
Does our equivalence suggest that the D1-instantons are not independent
of the Chern-Simons degrees of freedom?

\item {\bf Real versions.}
The framework first proposed by Berkovits \cite{Berkovits:2004hg} 
and the topological A-model of \cite{Neitzke:2004pf} seem to prefer
the real version of the twistor space, $\mathbb{RP}^{3|4}$, and correspondingly
real values of the moduli. Is there a real variation of our procedures?
One can imagine that the disconnected rules for the amplitudes might be derived
from the cubic twistorial string field theory of \cite{Berkovits:2004tx} if $K$
stringy propagators are expanded in component fields, so that the different parts
of the worldsheet become effectively disconnected.

\item {\bf Choice of prescriptions.}
According to our analysis, there is significant freedom to choose a 
twistor prescription for tree diagrams; we gave $d$ different rules,
all of which agree up to overall prefactors.  Is this all one can say, or
would a more sensitive study give more information
about which is the ``correct'' prescription?  Does this proliferation of
prescriptions persist beyond tree level?

\item {\bf Loops and higher genus.} 
We only studied tree diagrams, corresponding to genus zero curves.  
What are the exact rules and equivalences between various formulae for 
loop and nonplanar amplitudes?
Our analysis suggests that an investigation of possible degenerations of genus $g$ curves 
should be relevant for the understanding of loop diagrams in the twistor string.

\end{itemize}

\section*{Acknowledgements}
We are grateful to Michal Fabinger, Peter Svr\v{c}ek, Cumrun Vafa, Anastasia Volovich and 
Edward Witten for very useful discussions.
This work was conducted during the period S.G.\ served
as a Clay Mathematics Institute Long-Term Prize Fellow.
S.G. is also supported in part by RFBR grant 01-02-17488.
The work of L.M.\ was supported in part by Harvard DOE grant
DE-FG01-91ER40654 and the Harvard Society of Fellows.  The
work of A.N.\ was supported by NSF grants PHY-0255841 
and DMS-0244464.

\renewcommand{\baselinestretch}{1}
\small\normalsize

\bibliography{physics}

\end{document}